\documentclass[useAMS]{mn2e}

\usepackage{graphicx,color,here}
\usepackage{amssymb}
\usepackage{amsmath}

\author{R. P. Eatough$^{1,2}$\thanks{E-mail:
reatough@mpifr-bonn.mpg.de}, N. Molkenthin$^1$, M. Kramer$^{2,1}$,
A. Noutsos$^1$, M. J. Keith$^{3,1}$,\newauthor B. W. Stappers$^1$, and
A. G. Lyne$^1$.\\$^1$ Jodrell Bank Centre for Astrophysics, Alan
Turing Building, School of Physics and Astronomy, The University of
Manchester,\\ Manchester, M13 9PL, United Kingdom.\\$^2$
Max-Planck-Institut f\"ur Radioastronomie, Auf dem H\"ugel 69, 53121,
Bonn, Germany.  \\$^3$ Australia Telescope National Facility, CSIRO,
P.O. Box 76, Epping, NSW 1710, Australia.} \date{Accepted 2010 May 22. Received 2010 May 8; in original form 2010 March 25}
\title{Selection of radio pulsar candidates using artificial neural
networks}

\begin{document}

\maketitle

\vspace{3cm}
\begin{abstract}
Radio pulsar surveys are producing many more pulsar candidates than
can be inspected by human experts in a practical length of time.  Here
we present a technique to automatically identify credible pulsar
candidates from pulsar surveys using an artificial neural network. The
technique has been applied to candidates from a recent re-analysis of
the Parkes multi-beam pulsar survey resulting in the discovery of a
previously unidentified pulsar.
\end{abstract}


\pagenumbering{arabic}
\pagestyle{plain}

\section{Introduction}
Since the discovery of pulsars by Jocelyn Bell and Antony Hewish at
Cambridge in 1967 using a pen chart recorder (Hewish et
al. 1968\nocite{hbp+68}), pulsar searching has come a long way.
Modern pulsar surveys use high performance computing facilities to
perform an extensive range of signal processing and search algorithms.
These methods are designed to maximize sensitivity to weak, rapid, and
dispersed pulsar signals often buried in large amounts of terrestrial
radio frequency interference (RFI), or even in binary systems. There
is little doubt that these complex algorithms have aided searches for
pulsars, however there remain certain search tasks for which standard
computer programs are of little use. In particular, the final stage of
a pulsar search, the selection of credible pulsar candidates for
follow-up observations, which still remains a task for a human since
the decision is visual and based on a number of combined properties of
the pulsar signal. The process can be time consuming and inefficient
in analysis of large-scale surveys that produce many millions of
pulsar candidates.

Large-scale pulsar surveys such as the Parkes multi-beam pulsar survey
(PMPS) (Manchester et al. 2001)\nocite{mlc+01} have dramatically
increased the number of known pulsars. Finding more pulsars elucidates
the properties of their Galactic population, and also offers the
possibility of uncovering new and extreme phenomena in neutron-star
astrophysics. Future pulsar surveys will be done with the next
generation of radio telescopes, such as LOw Frequency ARray (LOFAR),
the Five hundred metre Aperture Spherical Telescope (FAST), and the
Square Kilometre Array (SKA) (e.g. van Leeuwen \& Stappers
2010\nocite{2010A&A...509A...7V}, Smits et al.,
2009a,b\nocite{2009A&A...505..919S}\nocite{2009A&A...493.1161S},
Cordes et al., 2004\nocite{2004NewAR..48.1413C}).  These radio
telescopes will be excellent survey tools because of their large
collecting areas, capability to form many simultaneous beams on the
sky, and in the case of the interferometers, wide fields of view.  It
is expected that these instruments will detect a large fraction of the
observable pulsars in the Galaxy. The inevitable flood of pulsar
candidates that will require inspection to achieve this will certainly
require some form of multi-person or machine based candidate
selection. Some large-scale astronomical surveys and data mining
projects have resorted to employing many online volunteers to search
for or classify their objects of interest in so-called `citizen
science' projects (e.g. Lintott et al.,
2008\nocite{2008MNRAS.389.1179L}, Westphal et al.,
2005\nocite{2005LPI....36.1908W}). In searches for pulsars, the pulsar
Arecibo L-band Feed Array (ALFA) survey collaboration has enlisted the
help of High School and undergraduate students in a successful
outreach and science program to identify potential pulsar candidates
(Jenet et al., 2007\nocite{2007AAS...211.0517J}).

Recent machine solutions include, candidate ranking based on
likelihoods calculated from parameter distributions of pulsar and
non-pulsar signals (Lee, private communication), and the sorting of
candidates based on a number of `scores' that indicate the similarity
of the signal to that of a typical pulsar (Keith et al.,
2009)\nocite{kel+09}. In this paper we present an alternative method
whereby an artificial neural network (ANN) has been trained using a
particular set of scores to automatically identify credible pulsar
candidates from a recent re-analysis of the PMPS. ANNs have long been
used in other areas of astronomy, for example in the morphological
classification of galaxies (e.g. Storrie-Lombardi et al.,
1992\nocite{1992MNRAS.259P...8S}, Zhang, Li \& Zhao
2009\nocite{2009MNRAS.392..233Z}), the estimation of photometric
redshifts of sources in the Sloan Digital Sky Survey (Firth, Lahav \&
Somerville 2003\nocite{2003MNRAS.339.1195F}) and in the selection of
microlensing events from large variability surveys
\nocite{2003MNRAS.341.1373B} (Belokurov, Evans \& Du 2003).

The outline of this paper is as follows: In Section 2 we describe how
pulsar candidate selection in our recent re-analysis of the PMPS is
typically done and some of the problems associated with this
method. Section 3 gives a brief introduction to ANNs. In Section 4 we
describe our implementation of an ANN to classify pulsar candidates
from our re-analysis of the PMPS and the results from a test on a
sample of the search output data.  Preliminary parameters of the
pulsar discovered using an ANN during our re-analysis of the PMPS are
given in Section 5. Finally Section 6 gives a summary, followed by a
discussion of the future application of ANNs to pulsar surveys.

\section{Candidate selection}
\begin{figure}
\begin{center}
\includegraphics[scale=0.335]{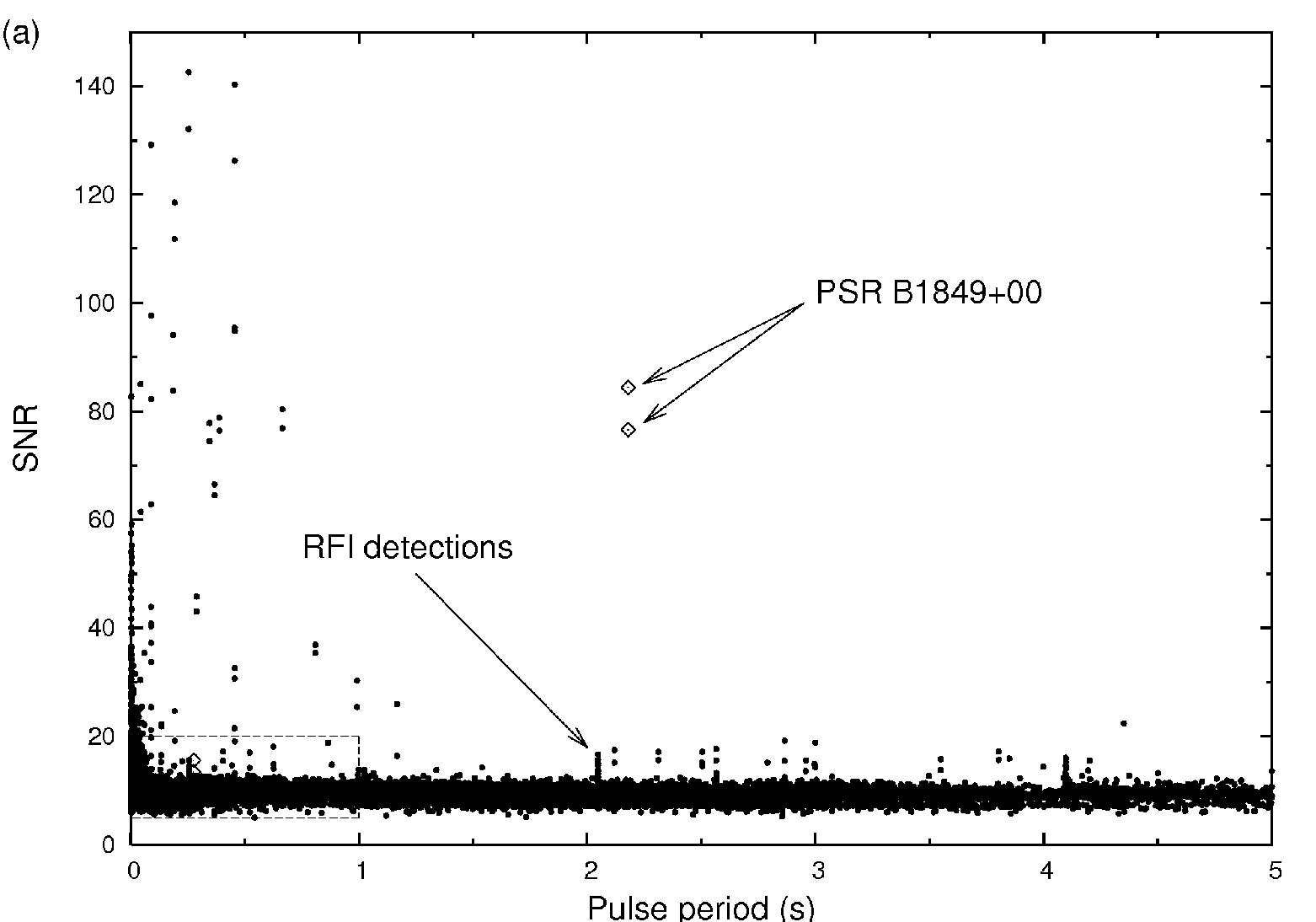}
\includegraphics[scale=0.335]{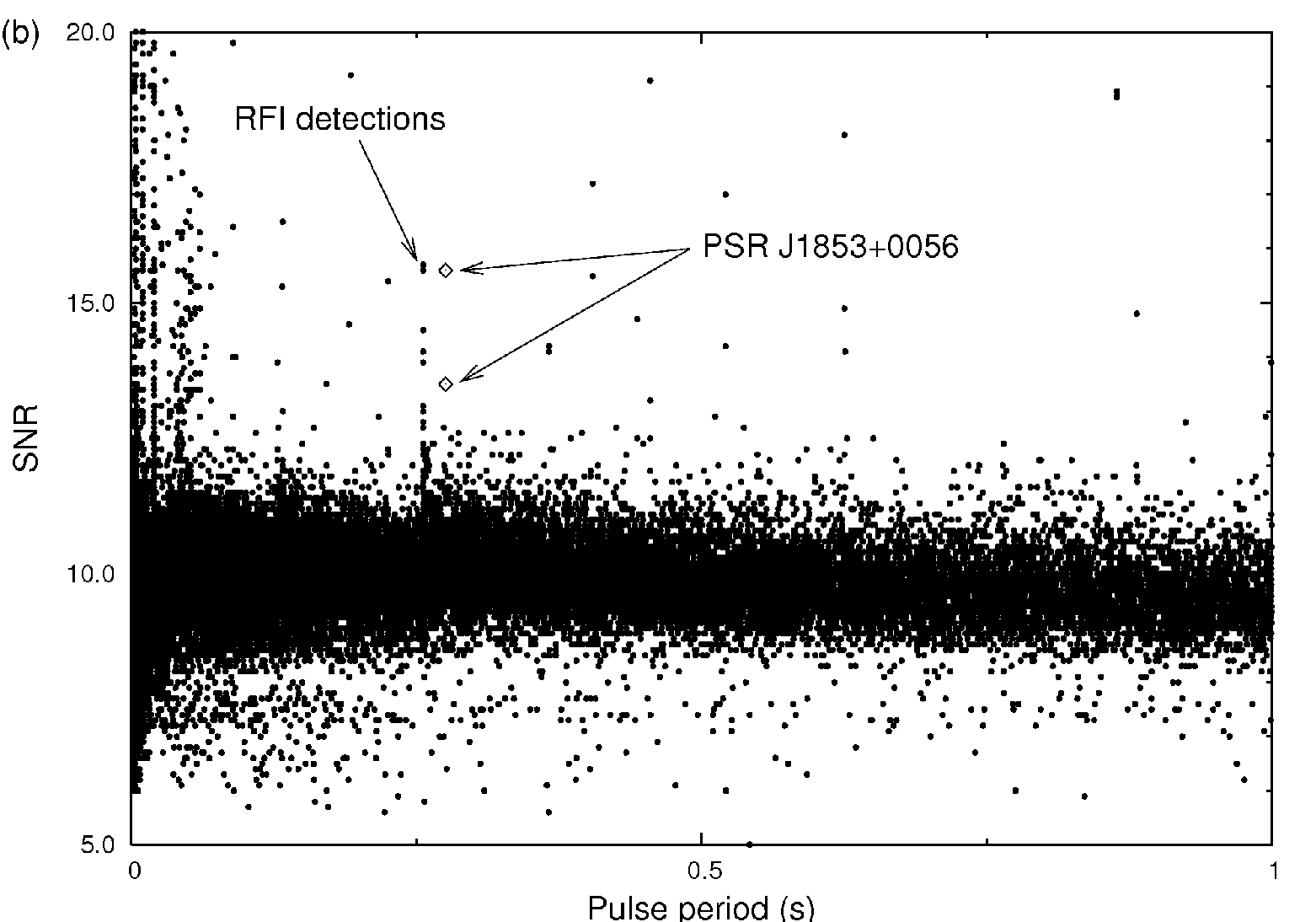} 
\caption{The barycentric pulse period-SNR distribution of candidates
from 650 survey beams of the PMPS.  The top panel (a) shows the entire
sample with two detections of PSR B1849+00 highlighted.  The area
enclosed by the box in the bottom left is shown in panel (b). Here the
position of two detections of PSR J1853+0056 have been marked.}
\label{fig:reaper}
\end{center}
\end{figure}

\begin{figure*}
\begin{center}
\includegraphics[scale=0.5]{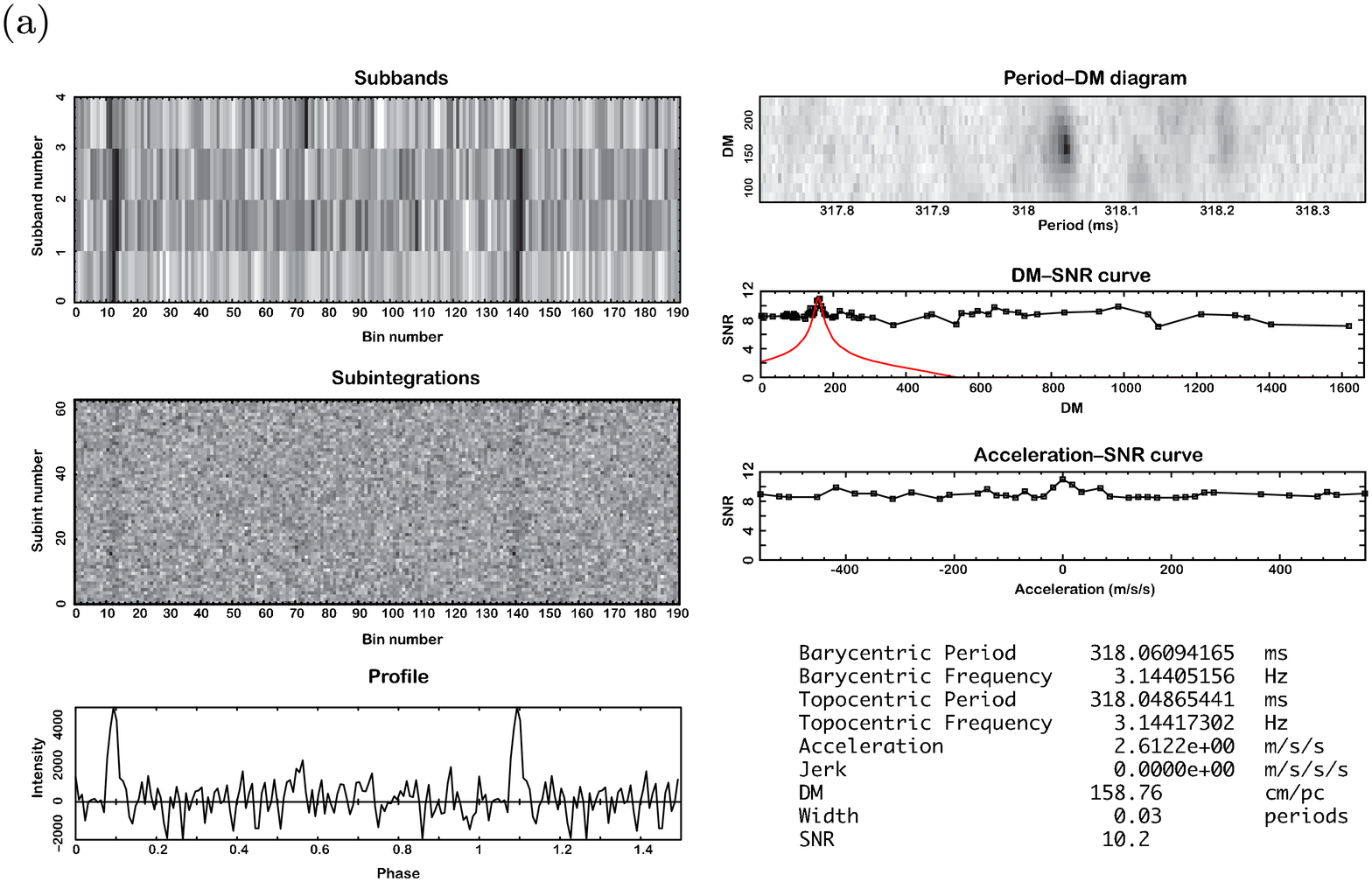}
\hspace{0.45cm}
\includegraphics[scale=0.5]{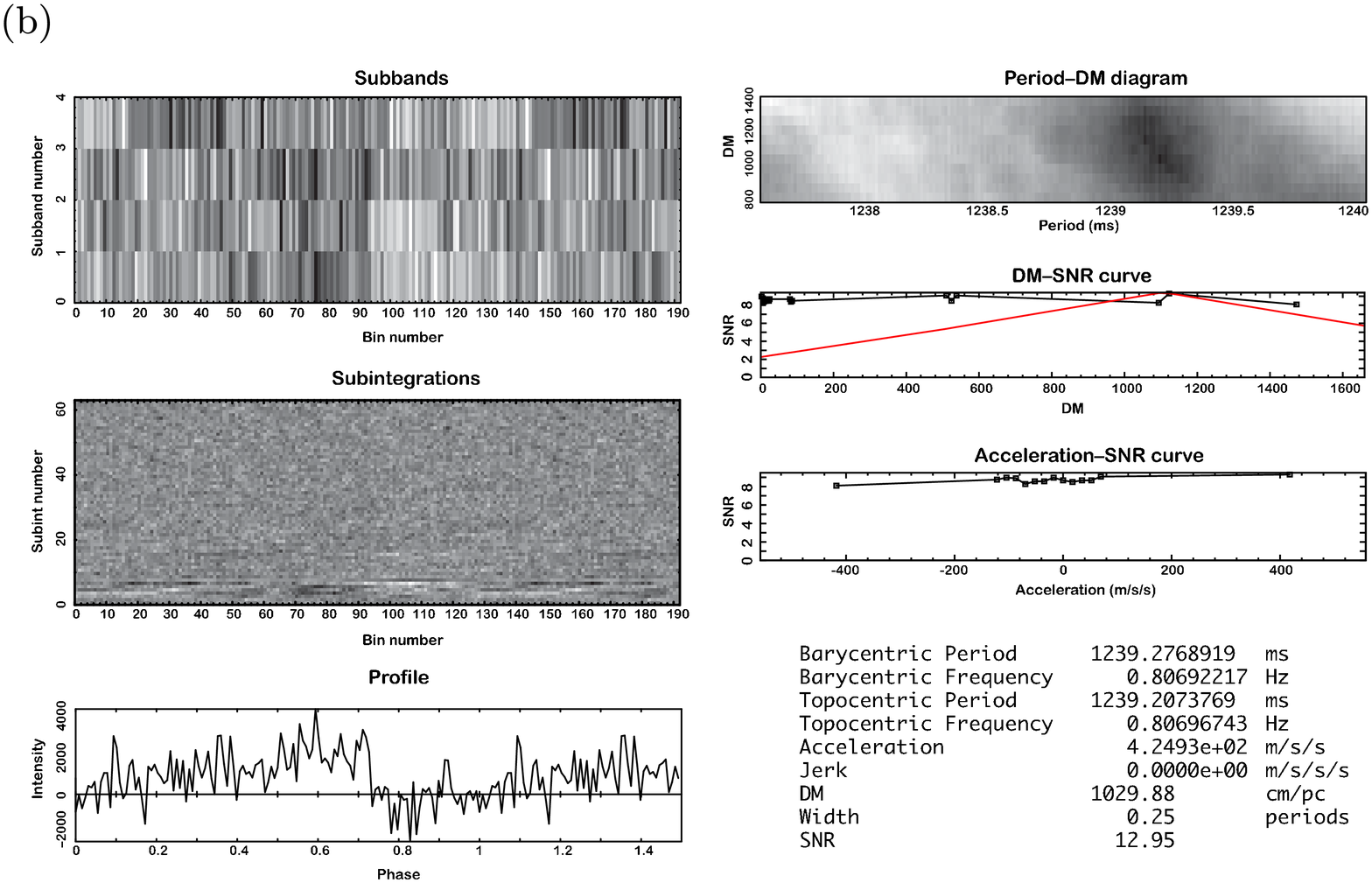}
\caption{Example pulsar candidate plots. Panel (a) shows PSR
J1926+0739, discovered in this work using an ANN (see
Section~\ref{s:disc}). Panel (b) shows a typical pulsar candidate that
is most likely the result of RFI. Starting clockwise from the bottom
left the two candidate plots display: the integrated pulse profile,
folded at the optimum period and dispersion measure (DM); 64 temporal
subintegrations of the observation, showing how the pulse varies with
time; stacked pulses across four frequency subbands, showing how the
pulse varies with observing frequency; the period-DM diagram, that
shows how the SNR varies with small changes in the folding period and
DM; the spectral SNR as a function of a wide range of trail DMs (the
DM-SNR curve), and finally the spectral SNR as a function of trial
acceleration (the acceleration-SNR curve).}
\label{fig:typ_cand}
\end{center}
\end{figure*}

The process of searching for undiscovered pulsars can be separated
into two distinct stages: firstly, the survey data are acquired at
radio telescopes and then processed using the standard search methods;
secondly, the output of the processing is analysed to select good
pulsar candidates for follow-up observation.  Detailed descriptions of
the steps involved in the first stage of a pulsar search can be found
in, for example, Lorimer \& Kramer (2005)\nocite{lk05}. The second
stage, which is the subject of this paper, is typically done by visual
inspection of the pulsar candidates generated from the initial
processing. Using a graphical selection tool, such as {\sc reaper}
(Faulkner et al. 2004\nocite{fsk+04}) or more recently {\sc jreaper}
(Keith et al. 2009\nocite{kel+09}), each pulsar candidate can be
presented as a point on a two-dimensional phase-space diagram like
those displayed in Figures~\ref{fig:reaper}(a) and
\ref{fig:reaper}(b). In this example, candidates from {650 survey
beams} of our latest re-analysis of the PMPS with acceleration
searches (Eatough et al., in prep) are plotted on axes of barycentric
pulse period versus signal-to-noise ratio (SNR). Large numbers of
spurious candidates due to RFI are visible as `columns' in this
phase-space. Selecting a point in the phase diagram would reveal the
so-called `candidate plot' for the pulsar candidate in question. Each
candidate plot contains a number of associated diagnostic plots
designed to help assess the credibility of pulsar candidates.
Examples of two such candidate plots can be seen in
Figures~\ref{fig:typ_cand}(a) and ~\ref{fig:typ_cand}(b).
Figure~\ref{fig:typ_cand}(a) shows the candidate plot of PSR
J1926$+$0739, discovered in this work using an ANN (see
Section~\ref{s:disc}). Figure~\ref{fig:typ_cand}(b) shows one of the
many spurious pulsar candidates that is most likely the result of
terrestrial RFI. To a pulsar astronomer searching this phase-space, or
indeed any individual with modest training, a distinction between
genuine or credible pulsar candidates and those due to RFI or random
statistical fluctuations can easily be made, making it possible to
pick out good pulsar candidates that should be marked for follow-up
observation. Common features of genuine pulsar candidates include: a
narrow pulse width, typically around five per cent of the pulse
period; a pulse present over all the subintegrations; pulse phase
coherent emission across all frequency subbands; an `island' in the
period-DM diagram; good agreement between the real and theoretical
DM-SNR curve, and a good agreement between the real and theoretical
acceleration-SNR curve. Millisecond and binary pulsars often exhibit
slight differences to normal pulsars in their candidate plots.  Again,
these differences are easily recognizable by a human expert. It should
be noted that exceptions to the above characteristics do occur and in
general each pulsar displays a unique signature in its candidate plot.

\subsection{The candidate selection problem}
Although graphical tools have reduced the number of candidate plots
that need to be inspected, by allowing RFI signals to be easily
avoided, there are still many more than can be effectively viewed by a
human in a reasonable length of time. This is especially true if the
more numerous, weaker candidates (where the majority of undiscovered
pulsars will be found) are probed. Typically our search algorithms
produce around 200 candidate pulsar plots per survey beam. In the PMPS
$\sim40\;000$ survey beams were performed. Using our current search
algorithms this gives a total of $\sim$ 8 million pulsar candidates
for visual inspection. In our search for fast binary pulsars the PMPS
beams have been split into two independent halves effectively doubling
this number. The typical time taken to inspect a candidate pulsar
depending on its credibility can lie between one and $\sim 300$ s. As
an example, a database of 1 million candidate pulsars would take
somewhere between $12$ days and $10$ years of continuous viewing to
search. In our search for highly accelerated pulsars the inspection
time is often over an hour as different acceleration algorithms and
integration lengths are used to find the optimal detection
SNR. Because viewing the pulsar candidates takes large amounts of
time, fatigue can also increase the potential for human error.

We wish to increase the speed and efficiency with which pulsar
candidates can be searched and also to avoid the arbitrary `cuts' that
graphical selection tools impose, such as ignoring all candidates near
RFI signals and those in the noise floor. One solution has been to
rank pulsar candidates not simply by their SNR but based on a number
of scores from the diagnostic plots that indicate the similarity of
the signal to that of a typical pulsar (Keith et al.,
2009)\nocite{kel+09}. This method has had a remarkable amount of
success, particularly at finding weaker pulsars that were missed in
previous analysis of the PMPS. However, the decision boundary on the
credibility of a pulsar candidate is complex and is based on the
combined features of each of the diagnostic plots and how they relate
to each other for the particular candidate in question. A simple
example of such a relation is that longer period pulsars, or those
with larger pulse widths, exhibit somewhat flatter DM-SNR curves. In
practice there are many more relations between the features displayed
in each of the diagnostic plots. At the moment scoring algorithms
cannot account for these relations and only provide features to weight
the individual scores.

Here we present an extension to the scoring method whereby an ANN has
been trained, using a set of scores, to attempt to model these
complex decisions and automatically identify credible pulsar
candidates. Firstly we give a brief introduction to ANNs.

\section{Artificial neural networks}
An ANN is a computational technique based loosely upon models of the
behavior of the human central nervous system. They have applications
to problems where some form of non-parametric estimation is required
i.e. where no known theory can be used to model the properties of a
set of data. Examples might include speech or pattern recognition. For
a more detailed description of ANNs and their applications see for
example Bishop (1995)\nocite{Bishop95}. A simplified picture based on
ANNs used in this work is as follows:

\begin{figure}
\begin{center}
\includegraphics[scale=0.5]{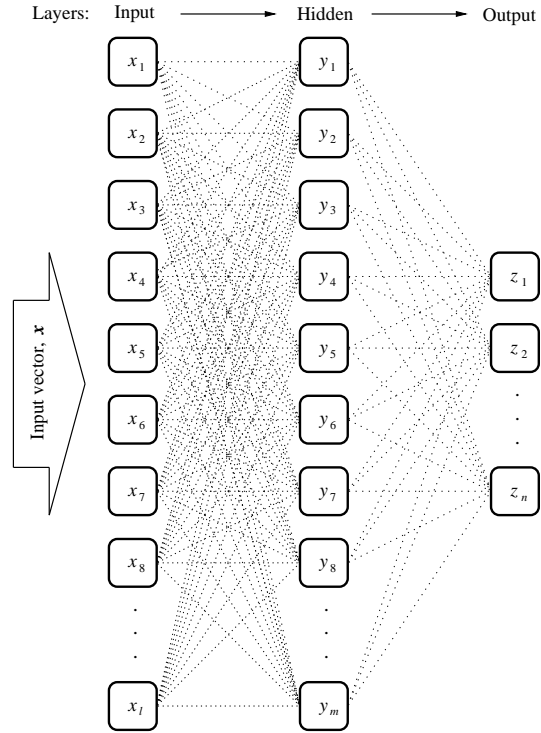}
\caption{A schematic diagram of a general three layered ANN of the
type used in this work.} 
\label{fig:architecture}
\end{center}
\end{figure}

\begin{figure}
\begin{center}
\includegraphics[angle=-90,scale=0.25]{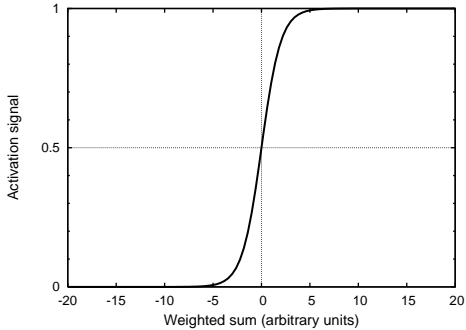}
\caption{An example sigmoid activation function. Using this function
the output of each node can be interpreted as a posterior
probability.}
\label{f:sigmoid}
\end{center}
\end{figure}

A number of nodes\footnote{\footnotesize Here the term `nodes'
represents individual processing elements of a computer program.} are
arranged into layers usually comprising an input layer, any number of
hidden layers, and an output layer. The architecture of an ANN can be
written simply by indicating the number of nodes in each of these
layers.  For example, one of the ANNs used in this work has twelve
nodes in the input layer, twelve nodes in one hidden layer, and two
nodes in the output layer giving the architecture 12:12:2. For
simplicity we now consider an architecture of just three layers with
any number of nodes, as depicted in Figure~\ref{fig:architecture}. The
input layer takes the `input vector', {\boldmath $x$}$=x_{\rm
1},x_{\rm 2},x_{\rm 3},...x_{l}$ a set of parameters to be passed to
the ANN. The hidden layer can also be written as a vector,
{\boldmath$y$}$=y_{\rm 1},y_{\rm 2},y_{\rm 3},...y_{m}$ and the output
layer by another, {\boldmath$z$}$=z_{\rm 1},z_{\rm 2},z_{\rm
3},...z_{n}$. The nodes in each layer are connected to all the nodes
in the previous layers. At each node in the hidden layer {\boldmath
$y$}, a linear weighted sum $S^y_j$ of the values from the previous
layer is calculated,
\begin{equation}
S^{y}_{j}=\sum\limits_{i=1}^{l}w_{ij}x_{i}
\end{equation}
where $i=1,2,3,...l$ in the {\boldmath $x$} layer, $j=1,2,3,...,m$ in
the {\boldmath $y$} layer, and $w_{ij}$ are weights associated between
nodes in each layer and which initially take random values. Positive weights
represent `affirmative' connections and negative weights represent
`inhibitory' connections. Next the so-called activation function, usually of
sigmoid form is calculated in the interval from zero to one (to
control the amplitude of the output of a node whatever the magnitude
of the weighted sum. See Figure~\ref{f:sigmoid}) and assigned to each
element in the {\boldmath $y$} layer,
\begin{equation}
y_{j}(S^{y}_{j})=1/[1+{\rm exp}(-S^{y}_{j})].
\end{equation}  
These values are then passed to the next layer (in our case the output
layer, {\boldmath$z$}) where another set of weighted sums,
$S^{z}_{k}=\sum\limits_{j=1}^{m}w_{jk}y_{j}$ are used to compute the
next set of activation functions, $z_{k}(S^{z}_{k})$ for
$k=1,2,3,...,n$. Because this is the last layer the values of these
functions constitute the output values of the ANN.

In a process of so-called `supervised learning' the ANN is trained by
comparison of the values from the output layer, {\boldmath$z$} with a
desired output vector, {\boldmath$d$} that is pre-determined by a
human . The comparison is usually done via the evaluation of a cost
function often called the `error', $E$. In `batch supervised learning'
the error function is averaged over many input vectors, i.e. a
training set, in one training iteration, $t$. Training is executed by
a process of modifying the weights in the next training cycle, $(t+1)$
to minimize the error function, $E$.  This is done in reverse starting
at the output layer, and ending up at the input layer, in the
so-called `backpropagation' phase:
\begin{equation}
w_{ij}(t+1)=w_{ij}(t)+\Delta w_{ij}(t+1),
\end{equation}
where $\Delta w_{ij}(t+1)$ are the sum of the weight modifications
between the input layer and the hidden layer in the $(t+1)$'th
cycle. 
Exactly how the weights are modified, i.e. the value of $\Delta
w_{ij}(t+1)$ depends upon the learning model chosen. In our system we
have chosen Resilient backPROPagation, (RPROP) where the size of
$\Delta w_{ij}(t+1)$ is fixed and depends only upon the sign of a
partial derivative, $\partial E/\partial w_{ij}$. For a detailed
description of the RPROP learning function see the Stuttgart Neural
Network Simulator (\textsc{snns}) user manual, version~4.2 and
references therein.\footnote{\footnotesize
http://www.ra.cs.uni-tuebingen.de/SNNS/} The RPROP
error function to be minimized is given by,
\begin{equation} 
E=\sum(z_{k}-d_{k})^2 + 10^{-\alpha} \sum w_{ij}^2 
\end{equation}
where $\alpha$ is an additional free parameter called the
`weight-decay' term. Once training is complete the weights are fixed
and the ANN can be fed input vectors with no training parameters. Any of
these vectors similar to those in the training set should produce output
vectors similar to the initial desired vector {\boldmath$d$}.

\section{Implementation in re-analysis of the PMPS}
The decision was taken to analyze some of the $\sim$ 16 million pulsar
candidates generated by our re-analysis of the PMPS using an ANN. All
ANNs presented in this work have been implemented using \textsc{snns}.

\subsection{Candidate scores}
As input to the ANNs small but characteristic input vectors were
formed from various candidate parameters and each of the panels
displayed in the candidate plots generated by our search software
(Figures~\ref{fig:typ_cand}(a) and ~\ref{fig:typ_cand}(b)).  Each
element of the input vector, which we term score, was designed to draw
a distinction between pulsar-like and non-pulsar-like signals, caused
by RFI or random statistical fluctuations (noise). In total twelve
scores, listed in Table 1, were used. Scores nine to twelve were
introduced to better represent the two-dimensional subband and
subintegration panels in the candidate plots. More details on the
features of candidate plots and scoring pulsar candidates can be found
in Keith et al. (2009)\nocite{kel+09}. Pre-processing of the candidate
plots in this way is beneficial both computationally and because of
the so-called `curse of dimensionality', which requires that the
larger the number of input parameters the bigger the training set
required to train the ANN effectively (see e.g. Bishop 1995, Section
1.4). This prevents the use of the raw data that make up the candidate
plot, for example the pulse profile, as an input.  With these simple
scores pre-processing of the candidates for the ANN took a small
fraction of the total data processing time.

A practical consideration worth mentioning at the pre-processing stage
is that candidate files should be in a format that can be readily
converted into input vectors for an ANN.

\subsection{ANN architectures}
Because of the large volume of data involved only two ANN
architectures have been investigated in this work. Firstly, a 8:8:2
followed by a 12:12:2 ANN. In general a large number of input
parameters is beneficial since the input data can be better
represented (up until the dimensionality problem takes effect). In
addition ANNs with increased complexity have more free parameters, in
the form of adjustable weights, and can in principle recreate more
complicated decisions. However, increased complexity comes at the cost
of computational efficiency. Large ANNs take longer to train and
longer to process input data. A good balance between these factors is
critical. In both ANNs investigated in this work each input vector has
an equivalent number of elements in a hidden layer followed by an
output layer comprising of two elements, representing probabilities of
the candidate being either a genuine pulsar or not a pulsar.

\subsection{Training}
\begin{figure}
\begin{center}
\includegraphics[scale=0.30,angle=-90]{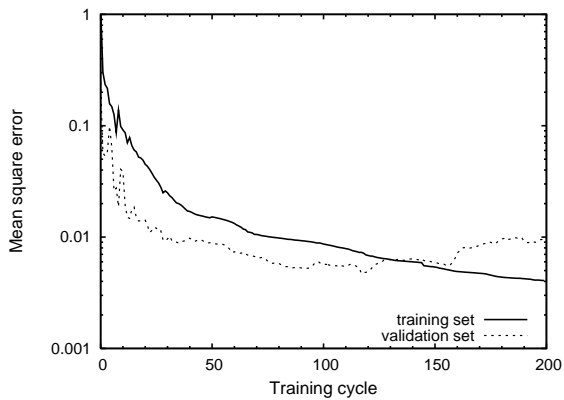}
\caption{The error development of the training and validation sets in
one of the ANNs investigated in this work. Training should be stopped
around the cycle corresponding to the minimum in the validation set.}
\label{fig:train}
\end{center}
\end{figure}

To form our training set 259 input vectors from a wide variety of
known pulsars were used in addition to the vectors generated from 1625
non-pulsar signals due to RFI or noise fluctuations. As mentioned in
Section 4.2 the output vector comprised of two elements, a genuine
pulsar identification probability between 0 and 1 and a non-pulsar
identification probability also between 0 and 1. During training these
outputs are pre-determined and form the desired output vector
{\boldmath $d$}.  For real pulsars, {\boldmath $d$}$\;= 1,0$ and for
RFI or noise signals, {\boldmath $d$}$\;= 0,1$. In accordance with ANN
training procedures a validation set of 28 different pulsars and 899
different RFI and noise signals was used. The development of the
error, $E$ of our training and validation set with training cycle for
one of our ANNs can be seen in Figure~\ref{fig:train}. One training
cycle constitutes the sum over all the vectors in the training
sample. Validation errors were also computed after every training
cycle. All weights are initially set to random values. Optimum ANN
training is achieved just before the error in the validation set
begins to increase (\textsc{snns} user manual, version~4.2), in this
case just before $\sim$ 120 cycles. If training is continued after
this point the ANN is said to be `overtrained' and will not perform
well on input vectors that do not match those in the training set
precisely.

\begin{table}
  \begin{center}
    \caption{Summary of the scores used in the input vectors in this
       work. The `optimized' curves refer to smaller finer
       searches in DM and acceleration around the initial detection
       values from the wider searches. Scores marked with a
       $^{\clubsuit}$ are additions only included in the 12:12:2 ANN.}
      \begin{tabular}{|r|l|}
        \hline\hline
        \multicolumn{1}{c}{No.} & \multicolumn{1}{l}{Score description}\\\hline
	1 & Pulse profile SNR.\\
	2 & Pulse profile width.\\
	3 & Chi-square of fit to theoretical DM-SNR curve.\\
	4 & No. of DM trials with SNR $>$ 10.\\
	5 & Chi-square of fit to optimized theoretical \\
           & DM-SNR curve.\\
	6 & Chi-square of fit to theoretical\\ 
           & acceleration-SNR curve.\\
	7 & No. of acceleration trials with SNR $>$ 10.\\
	8 & Chi-square of fit to optimized theoretical\\ 
        &     acceleration-SNR curve. \\
	9 & RMS scatter in subband maxima.$^{\clubsuit}$\\
	10 & Linear correlation across subbands.$^{\clubsuit}$\\
	11 & RMS scatter in subintegration maxima.$^{\clubsuit}$\\
	12 & Linear correlation across subintegrations.$^{\clubsuit}$\\\hline
     \end{tabular}
  \end{center}
\end{table}

\begin{figure*}
\begin{center}
\includegraphics[scale=0.32,angle=-90]{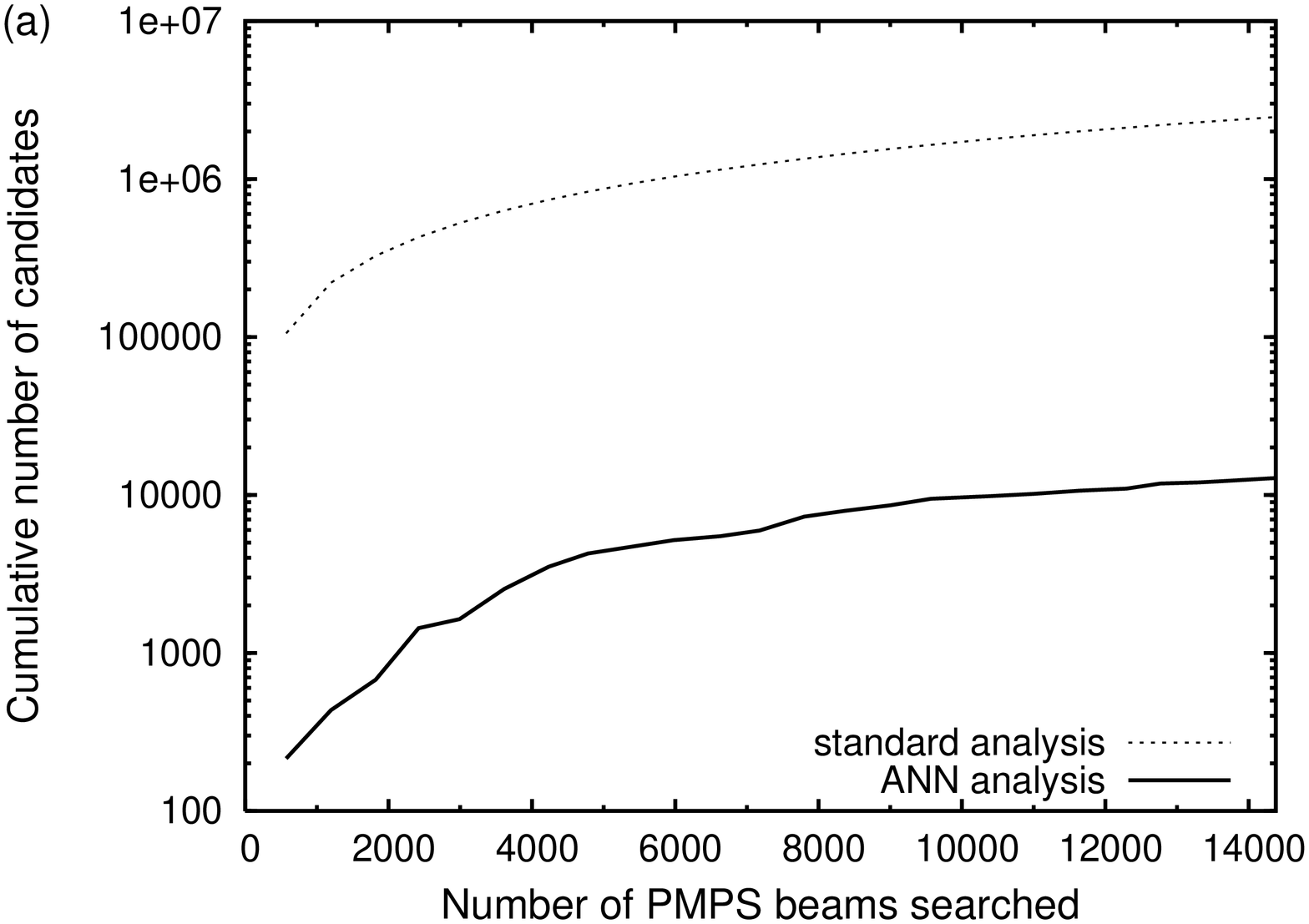} 
\hspace{0.45cm}
\includegraphics[scale=0.32,angle=-90]{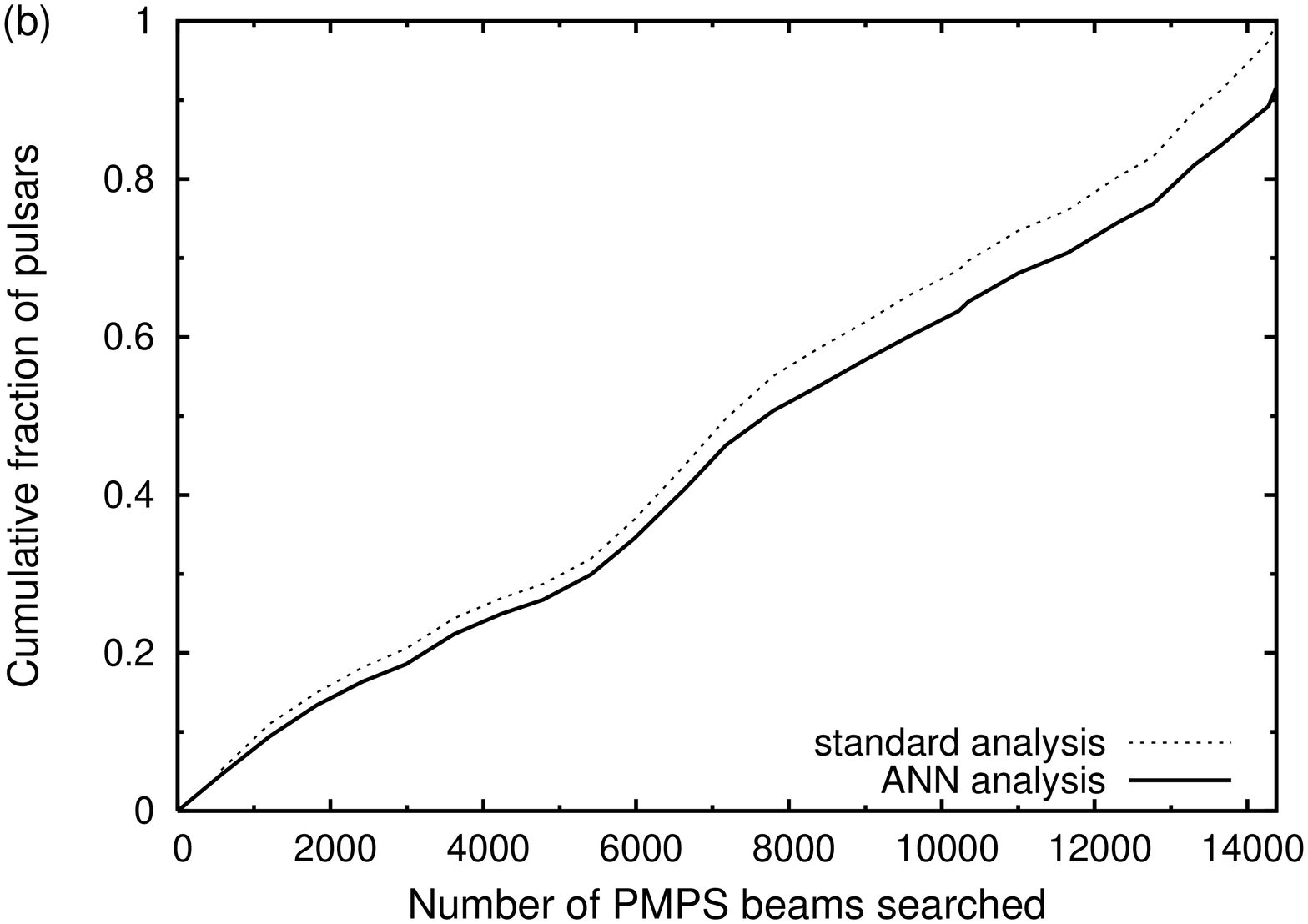}
\caption{Panel (a) shows the cumulative number of candidates recovered
by the standard analysis and the ANN over the results from $\sim
14\;400$ survey beams of our re-analysis of the PMPS. Panel (b) shows
the cumulative fraction of known pulsars in the sample recovered by
each method.}
\label{fig:pulsars_recovered}
\end{center}
\end{figure*}

\subsection{Test results}
\label{s:tests}
The 8:8:2 ANN has been tested on search results from a random
selection of $\sim14\;400$ survey beams from our re-analysis of the
PMPS.  Figure~\ref{fig:pulsars_recovered} summarizes the results of
both the standard graphical analysis and the analysis using the
ANN. In Figure~\ref{fig:pulsars_recovered}(a) the cumulative number of
candidates that would need inspection in both a standard analysis
using a graphical selection tool (neglecting candidates that are
typically avoided) and using the ANN is plotted against the number of
PMPS beams searched. Figure~\ref{fig:pulsars_recovered}(b) displays
the cumulative fraction of pulsars detected by each method. Our
standard analysis shows there are 501 detections of known pulsars in
$\sim2.5$ million pulsar candidates from the $\sim14\;400$ beams. Due
to constraints on the available amount of processed data 51 of the 501
pulsars in the test sample were also present in the training set.
Ideally the test sample should be entirely independent of the training
set to test if the ANN has been overtrained. In the ANN analysis a
candidate was flagged for viewing if the output vector of the ANN,
{\boldmath $z$}$\;=z_{1},z_{2}$, had values $z_{1}> 0.5$ and
$z_{2}<0.5$, otherwise it was ignored.  The number of candidates that
filled this criterion was low, of the order $0.5$ per cent
corresponding to around $13\;000$ candidates. Encouragingly from the
$\sim 13\;000$ candidates selected by the ANN $92 $ per cent of the
pulsars present in the test sample have been recovered. This gives the
statistics that 1 in every $\sim 30$ candidates selected by the ANN is
a pulsar whereas only 1 in every $\sim 4900$ candidates that would be
viewed in a standard graphical analysis will be a real pulsar
detection. Practical use of graphical selection tools means that not
all the candidates would need to be viewed however this increases the
risk of missing weak pulsars or those with similar periods to RFI
signals. Ignoring the training pulsars that were present in the test
sample reduces the recovered fraction of pulsars by $1$ per cent.

The addition of four more scores in the 12:12:2 ANN marginally
improved the recovered fraction of pulsars to $93 $ per cent. However the
recovered number of candidates, along with the computation time,
nearly doubled. This small improvement was somewhat expected. The
relatively low resolution of the two-dimensional diagnostic plots in
our re-analysis of the PMPS make spotting statistical trends in these
planes difficult. In the candidate plots from current and future
pulsar surveys these planes will be better sampled due to the increase
in available computational power.

\begin{figure*}
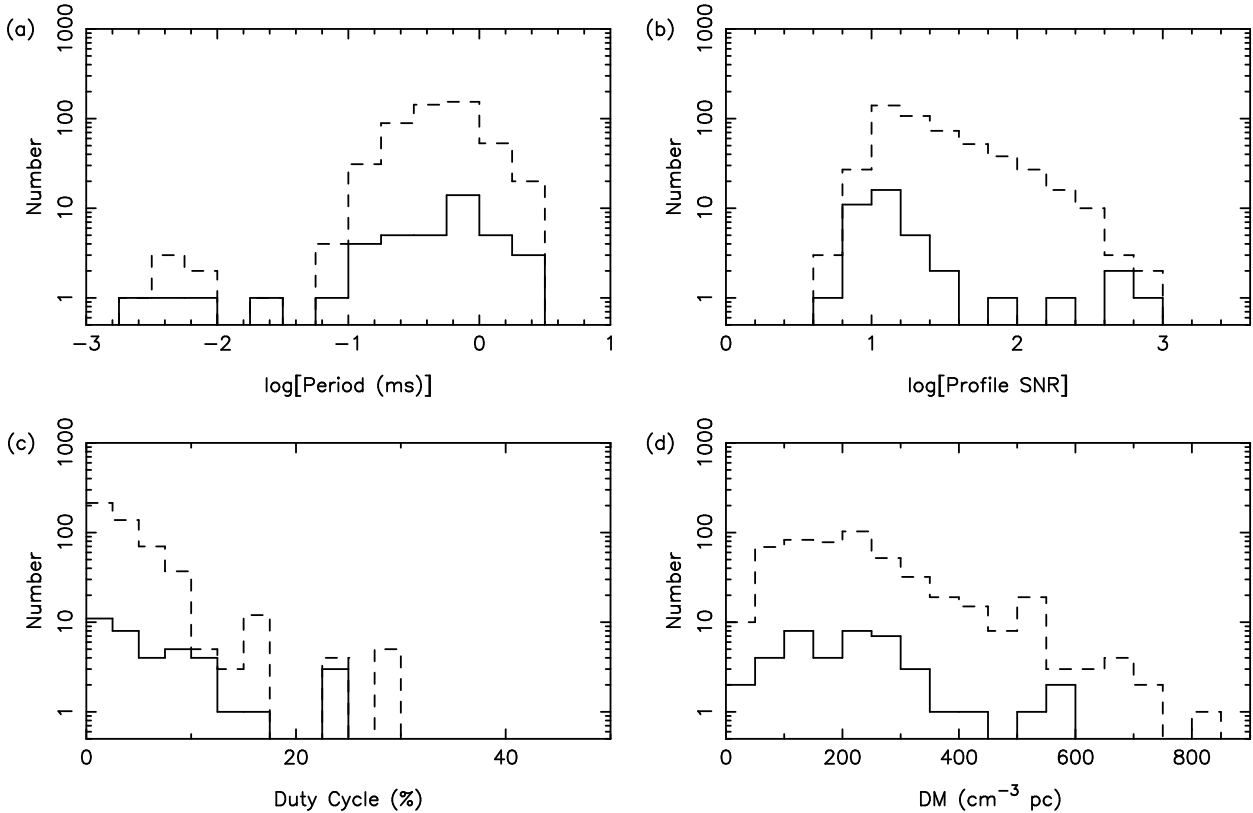

\centering
\begin{tabular}{cc}
\includegraphics[scale=0.32,angle=-90]{figs/pers_final.ps} &
\includegraphics[scale=0.32,angle=-90]{figs/snrs_final.ps} \\ 
\includegraphics[scale=0.32,angle=-90]{figs/duty_final.ps} &
\includegraphics[scale=0.32,angle=-90]{figs/dms_final.ps}  \\
\end{tabular}
\caption{Histograms showing the distributions of (a) spin period, (b)
pulse profile SNR, (c) pulse duty cycle, and (d) DM values of all the
pulsars in the test sample (dashed lines) and pulsars missed by the ANN
(solid lines).}
\label{fig:props}
\end{figure*}

The properties of the pulsars missed by the ANN have been
investigated. Distributions of spin period, pulse profile SNR, pulse
duty cycle and DM for pulsars in the sample and those missed by the
ANN are displayed in Figure~\ref{fig:props}.  There is no significant
correlation between those pulsars missed and any property
investigated. Two features of the distributions of missed pulsars are
however worth noting. Firstly, around $50$ per cent of the pulsars
with periods less than 10~ms were missed by the ANN. This result was
somewhat expected and can be explained by the different appearance of
the candidate plots associated with millisecond pulsars (MSPs). For
example, MSPs can appear to have larger pulse widths because of the
decreased number of phase bins across the pulse profile. Also they
exhibit narrow DM-SNR curves due to the rapid drop off in sensitivity
with incorrect DM trial. Unfortunately only 6 MSP detections were
present in the test sample making this result ambiguous.  Secondly, 60
per cent of the pulsars with the profile SNRs $\gtrsim 400$ are
missing. This has been shown to be due to abnormal DM-SNR curves
produced by our search software for extremely bright pulsars.  The
training set somewhat overrepresented candidates from RFI or noise
signals. This was done with the aim of identifying these more numerous
candidates effectively, however it could also have increased the
number of false positives for this particular class. As a result the
other missing pulsars may simply have been misidentified.

The analysis of the $\sim2.5$ million pulsar candidates using the
8:8:2 ANN took a total of $\sim50$ CPU hours. Since the process is
fully `parallelisable' the task was easily distributed across a
computer cluster.

\section{Discovery of a pulsar using an ANN} 
\label{s:disc}
The ANN has been applied to a portion of candidates generated from our
re-analysis of the PMPS with acceleration searches. In total 14
pulsars have been discovered using standard candidate selection
techniques (Eatough et al. 2009\nocite{ekl+09}, Eatough et al. in
prep), and one has been found using the 8:8:2 ANN (see
Figure~\ref{fig:typ_cand}(a)). The pulsar, of provisional name PSR
J1926$+$0739, has a spin period of $\sim \,$318 ms, a DM$\, \sim
\,$160~${\rm cm^{-3}pc}$ and was detected with SNR$\, \sim \,$10.
This pulsar is now being timed at the Jodrell Bank Observatory and
will be presented in (Eatough et al. in prep) when the timing solution
is complete and all astrometric and spin-down parameters have been
determined with sufficient accuracy.

\section{Summary \& Discussion}
ANNs have been developed to aid searches for radio pulsars in a
database of $\sim$ 16 million pulsar candidates generated from a
recent re-analysis of the PMPS. The ANNs were trained using small sets
of characteristic scores derived from the pulsar candidate plots
generated for human inspection. Around $92$ per cent of the pulsars
present in a test sample of $\sim$ 2.5 million pulsar candidates can
be recovered by our ANNs. 

It is likely that pulsars from the test sample were missed due to one
of three reasons: poor training of the ANNs on MSPs, abnormal
candidate plots generated by our search software, or unbalanced
training sets.  Future implementations should avoid unbalanced
training sets or bias the training sets toward the objects of
interest. A larger number of false positives from the ANN is
beneficial for searches of the inherently rare objects in the data set
viz. pulsars \nocite{2003MNRAS.341.1373B} (e.g. Belokurov, Evans \& Du
2003).

To test if the `contamination' from non-pulsars in the results set
could be further reduced the threshold for pulsar identification,
$z_{1}$, was increased to $0.7$, $0.8$, and $0.9$. In each case the
number of genuine pulsar to non-pulsar candidates recovered by the ANN
were 1:24, 1:23, and 1:20 respectively, however the corresponding
recovered fraction of pulsars were 89, 86, and 81 per cent. In a
search for pulsars these few per cent could be potential new
discoveries. As such higher thresholds should only be used when the
number of candidates produced by the ANN are unmanageable. Improving
the detection efficiency of our ANNs will be the subject of future
work. Improvements in this area are likely to depend upon better
representation of the two-dimensional diagnostic plots in our input
vectors. These planes are particularly useful for discriminating
against narrow-band and impulsive RFI.

ANNs are only capable of classifying the inputs given to them based on
the training received. By training the ANN using a set of pulsars with
particular characteristics we reduce the possibility of uncovering
atypical, unusual, or unexpected phenomena. Therefore, it is important
to use training sets of pulsars with a wide variety of properties and
displaying phenomena such as scintillation, scattering, intermittency
and binary motion. Without training on pulsars that exhibit such
properties an ANN will only select against these objects. As described
in Section~\ref{s:tests} MSPs display somewhat different candidate
plots to their standard pulsar counterparts. Separate ANN training
using only data from MSPs may be required to find these pulsars. For
more exotic pulsars, such as relativistic binary pulsars, any training
sample will be limited in size. To find these systems it might be
possible to form training sets from simulated pulsar signals covering
a wide range of pulsar and binary parameters. However, it is expected
that ANNs trained with simulated training sets will not perform as
well as those based on real data due to the subtle effects of
instrumentation and RFI environments. In new pulsar surveys where the
search data may vary from that of previous surveys, for example in the
number of frequency channels and the sampling time, dedicated
observations of known pulsars will be required to build a training
database of MSPs and the standard population of pulsars.

A complementary class of ANNs are unsupervised ANNs which require no
desired or `target' vector during training. The best example of such a
ANN is the Kohonen Self Organizing Map (SOM) (e.g. Kohonen,
2001)\nocite{Kohonen01}. SOMs provide a way of representing
multi-dimensional data in a lower number of dimensions, generally the
two-dimensions of a map. Similar input vectors share similar regions
on these maps. Such a ANN could be used to divide the pulsar
candidates into sub-classes such as MSPs, standard pulsars, noise
signals, and even common RFI signals. Although not optimal for
candidate selection such ANNs will be investigated in future work.

Future pulsar surveys will probe larger volumes of both real space and
the parameter space associated with radio pulsars. The SKA will have a
sensitivity that will allow the discovery of all $20\;000$ to
$30\;000$ pulsars beaming towards Earth and visible from the short
listed sites in South Africa and Western Australia. Inspecting the
candidates from such surveys will be a tremendous data analysis
task. For illustration, scaling the PMPS up to an all sky survey would
result in approximately a factor of 40 increase in the number of
pulsar candidates. Using the same search algorithm as in our recent
re-analysis of the PMPS and assuming an inspection time of one second
per pulsar candidate, viewing all the candidates from such as survey
would take over 20 years. This figure does not take into account beams
that contribute many erroneous candidates due to RFI and which are
relatively easy to flag and avoid using graphical selection tools.
However, as we have already discussed these cuts can increase the risk
of missing genuine pulsars that lie near to RFI in phase-space. Using
an ANN and assuming the same recovered fraction of candidates of $0.5$
per cent the total inspection time would come down to just over one
month, vastly reducing the work load on human observers.

We would like to stress that simple ANNs, like those presented in this
work, should not yet be treated as a complete replacement for human
inspection of pulsar candidates. Humans are still best placed to spot
anything unusual or interesting about the individual pulsar candidates
being viewed.  Because of the significant time savings, of around two
orders of magnitude, ANNs might be best used for first-passes over
search output data before a more detailed manual inspection with
graphical selection tools.

\section*{ACKNOWLEDGMENTS}
This research was partly funded by grants from the Science \&
Technology Facilities Council. The Australia Telescope is funded by
the Commonwealth of Australia for operation as a National Facility
managed by the CSIRO. N. Molkenthin acknowledges the support of the
Studienstiftung des deutschen Volkes and the ERASMUS exchange
programme. We thank John Brooke and members of The University of
Manchester School of Computer Science for useful discussions. We would
like to thank Sir Francis Graham-Smith, for useful discussions and
manuscript reading. We also acknowledge Alessandra Forti and The
University of Manchester Particle Physics group for use of the TIER2
computing facility\footnote {\footnotesize
http://www.hep.manchester.ac.uk/computing/tier2}.

\bibliographystyle{mnras}

\end{document}